# High-frequency magnetic response measurement of test mass with a fluxgate magnetometer for gravitational wave detection

Yuanyang Yu,[1] Butian Zhang,[1,*] Shengxin Lin,[2] Jianping Liang,[1] Donghua Pan,[2] Shun Wang,[1,†] and Ze-Bing Zhou[1]

[1]MOE Key Laboratory of Fundamental Physical Quantities Measurement and Hubei Key Laboratory of Gravitation and Quantum Physics, PGMF and School of Physics, Huazhong University of Science and Technology, Wuhan 430074, China

[2]Laboratory for Space Environment and Physical Science, Harbin Institute of Technology, Harbin, 150001, China

## Abstract

For space-borne gravitational wave detectors, such as LISA and TianQin, the disturbance caused by the coupling of test masses and the external magnetic fields is one of the main sources of the residual acceleration noise. Although the detection frequency band is from 0.1 mHz to 1 Hz, magnetic fields with frequencies higher than 1 Hz can still contribute to the noise through down conversion effect. Therefore, it is necessary to measure the AC magnetic susceptibility or magnetic response of the test mass at higher frequency for the evaluation of the magnetic noise. In this work, we propose a magnetic field response measurement method by directly probing the induced magnetic field of the test mass placed in a spatially uniform magnetic field. The frequency can be measured up to 1500 Hz, satisfying the requirement of space-borne gravitational wave detection.

---

* butianzhang@hust.edu.cn

† shun@hust.edu.cn

# I. INTRODUCTION

Space-borne gravitational wave detectors, such as LISA, TianQin and Taiji, typically consist of three spacecrafts forming an equilateral triangle. [1~3] The signal of gravitational waves is detected by measuring the change of the relative distance between test masses in free fall state inside the spacecrafts by laser interferometry. For TianQin, the residual acceleration of the test masses is required to be on the order of $10^{-15}\ \mathrm{ms^{-2}Hz^{-1/2}}$ in the measurement band from 0.1 mHz to 1 Hz. [2~6] In order to meet such stringent requirements, it is necessary to suppress the noise from various sources, including temperature variation, electric field and magnetic field. [7] The magnetic noise is one of the main sources of residual acceleration noise, which arises from the coupling between test masses and the external magnetic fields of the space and satellite. [8~9] For example, the magnetic noise accounts for 24% of total noise budget of TianQin. [2]

The magnetic forces are generated from the interaction of the remanent $m_r$ and induced magnetic moment $m_i$ of the test mass with the external magnetic field $B$ through $\nabla[(m_r + m_i)B]$. The remanent $m_r$ is generally independent of the applied magnetic field, while the induced magnetic moment $m_i$ represents the response of the test mass to the applied magnetic field, which can be divided into DC response and AC response. [10] The DC response is obtained by measuring $m_i$ of the test mass in the static magnetic field. In this case $m_i$ is equal to $\chi V \mu_0^{-1} B$, where $\chi$ is the DC volume magnetic susceptibility, $V$ is the volume of the test mass, $\mu_0$ is the permeability of vacuum. The DC magnetic susceptibility $\chi_0$ of the test mass, which is an Au-Pt alloy cube, mainly depends on its composition and manufacturing process. [11,12] The AC response of the test mass, requires measuring $m_i$ under an AC magnetic field. However, the origin of the AC response of the test mass is intricate, and a unified explanation has not yet been formulated. Nevertheless, since our primary concern is assessing the magnetic forces imposed by the AC magnetic field on the test mass, the focus of this work lies in measuring $m_i(\omega)$ under AC magnetic field $B(\omega)$, where $\omega$ is the frequency of the AC magnetic field. If define the response of the test mass to the AC magnetic field as $R(\omega) = \mu_0 m_i(\omega)/B(\omega)$ with $\nabla R(\omega) = 0$, then the test mass produces magnetic force $\nabla[m_i(\omega)B(\omega)]$ due to AC magnetic response through $R(\omega)\nabla[B^2(\omega)]$. Therefore, in order to accurately evaluate the magnetic force, an accurate measurement of $R(\omega)$ is necessary.

Under low-frequency magnetic fields, the diamagnetic and paramagnetic properties are the main sources of magnetic response of the test mass. The contribution of magnetic hysteresis loss such as eddy current to magnetic response grows with increasing frequency. [13~15] Under high-frequency magnetic fields, the eddy current effect becomes dominant. [16] The magnetic fields with frequency higher than 1 Hz, even though not within the detection frequency band of 0.1 mHz $\sim$ 1 Hz, can still cause noise through down conversion effect. According to LISA Pathfinder's recent results, two high-frequency magnetic fields with close frequencies $B_I(\omega + \Delta\omega)$ and $B_F(\omega - \Delta\omega)$ generate a low frequencies magnetic force $f(2\Delta\omega)$ by $R(\omega)\nabla(B_I \cdot B_F)$ on the test mass. [17,19] Therefore, it is crucial to measurement the $R(\omega)$ of the test mass under high frequencies magnetic field in order to evaluate the magnetic noise for gravitational wave detection. It should be noted that due to the complexity of the space magnetic field and the spacecraft magnetic field, [8] measuring the AC magnetic response cannot precisely assess the magnetic force generated during the test mass of the on-orbit state. Nevertheless, the evaluation of the AC magnetic response remains a key and versatile method for characterizing the magnetic properties of test mass affected by relatively uniform time-varying magnetic fields. This indicator plays an important role in providing an approximation of the magnetic effect.

To measure $R(\omega)$, we need to generate an approximately uniform magnetic field $H(\omega)$ and measure the induced magnetic moment $m_i(\omega)$ generated by the sample in the magnetic field. There are two kinds of methods for measuring the sample's induced magnetic moment $\boldsymbol{m_i}(\omega)$. The first method, such as ACMS (AC Measurement System), uses pick-up coils, or flux sensor (such as superconducting quantum interference device (SQUID) magnetometers) to detect the induced magnetic moment by measuring the change of magnetic flux. [20~23] However, such commercially available equipment is designed for measuring mm-sized samples, not satisfying the measurement requirements for cm-sized samples, such as the test mass. Challenges in the field uniformity, resolution and accuracy need to be overcome if one plans to use ACMS to measure larger samples.

The second method involve convert the magnetic moment into force or torque for indirect measurement, such as torsion balance or pendulum, [16, 24~26] which can measure cm-sized objects. Xu et al. measured the AC magnetic response of a 2-cm-side cubical-shaped copper sample using a torsion balance with dual-frequency modulation technique. [16] However, the torsion balance's measurement process is relatively complicated and long measurement time is needed. For example, the suspension of the test mass in the vacuum chamber is required and the measurement usually take days. Meanwhile, the highest measurement frequency is limited to less than 100 Hz. In order to compensate for these shortcomings, in this work, we designed a facile measurement scheme for AC magnetic response for cm-sized sample with high precision based on a magnetic fluxgate magnetometer. Although this method is based on directly probing weak induced magnetic fields, the experimental setup is carefully designed for ambient conditions such that no additional magnetic shielding is required.

## II. PRINCIPLE OF THE INSTRUMENT

The magnetic field at point $P$ generated by test mass can be expressed as:

$$\boldsymbol{B}(\boldsymbol{r}') = \frac{\mu_0}{4\pi}\iiint_V \left[\frac{3\cdot[\boldsymbol{m}(\boldsymbol{r})\cdot(\boldsymbol{r}-\boldsymbol{r}')]\cdot(\boldsymbol{r}-\boldsymbol{r}')}{(\boldsymbol{r}-\boldsymbol{r}')^5} - \frac{\boldsymbol{m}(\boldsymbol{r})}{(\boldsymbol{r}-\boldsymbol{r}')^3}\right] dV \tag{1}$$

Where $\boldsymbol{m}(\boldsymbol{r})$ is the magnetic moment per unit volume and the integration is over the whole volume of the test mass, $\boldsymbol{r}$ and $\boldsymbol{r}'$ are the position vector of point $P$ and $\boldsymbol{m}(\boldsymbol{r})$. Under the assumption that the magnetic moment is uniformly distributed, $\boldsymbol{m}(\boldsymbol{r})$ is constant, and Eq. (1) can be rewritten as:

$$\begin{bmatrix} B_x \\ B_y \\ B_z \end{bmatrix} = \begin{bmatrix} K_{xx} & K_{xy} & K_{xz} \\ K_{yx} & K_{yy} & K_{yz} \\ K_{zx} & K_{zy} & K_{zz} \end{bmatrix} \begin{bmatrix} M_x \\ M_y \\ M_z \end{bmatrix} = \boldsymbol{KM} \tag{2}$$

Where $\boldsymbol{K}$ is a position matrix, which is only related to the shape of the sample and the relative position between point $P$ and sample, and $\boldsymbol{M}$ is a vector, which represents the total magnetic moment of the sample. However, in an AC magnetic field, $\boldsymbol{m}(\boldsymbol{r})$ becomes a function of both the position $\boldsymbol{r}$ and frequency $\omega$ due to the existence of skin effect. [13] In this case, $\boldsymbol{m}(\boldsymbol{r})$ is no longer spatially uniform and the total magnetic moment of the sample is:

$$\boldsymbol{M} = \iiint_V \boldsymbol{m}(\boldsymbol{r}, \omega) dV \tag{3}$$

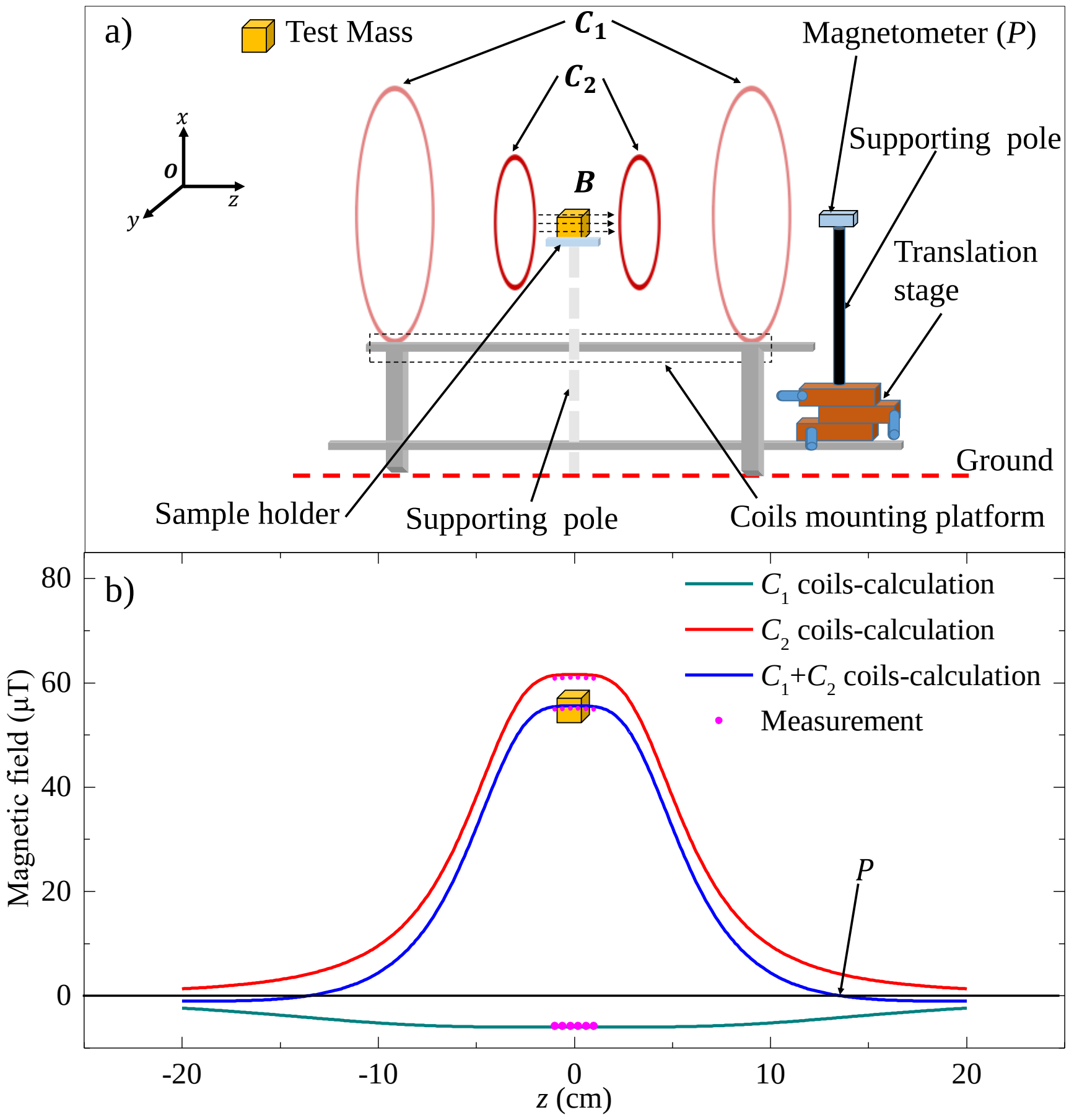

**Fig. 1. a):** Schematic diagram of the device. $C_1$ and $C_2$ are two pairs of Helmholtz coils of different radius that coincide in center and axial direction. **b):** The magnetic field in the $z$ direction (coincident with the coils axis) generated by coils changes with the $z$ direction coordinate. The center points of the two pairs of coils coincide, and the coordinate origin is located at the center point of the coils. The solid line is the theoretically calculated distribution of the coils magnetic field. The red dot is the distribution of the magnetic field at different positions near the origin measured by the magnetometer. The error between calculation and measurement is less than 1%.

This makes accurate determination of $\boldsymbol{M}$ difficult. This problem can be solved by placing the sample far away from the magnetometer. When the distance from the magnetometer to the center of the sample $D$ is much larger than the side length of the sample $L$, the sample can be treated as an effective magnetic dipole located at its center and the measurement error caused by the magnetic moment inhomogeneity of the sample can be ignored. Therefore, we used a fluxgate magnetometer located at distance far enough from the sample to measure the induced magnetic field.

The experimental apparatus is shown in Fig. 1(a). The magnetic field measured by the magnetometer mainly consists of two parts: $B_{zr}(\omega)$ and $B_{z\chi}(\omega)$, where $B_{zr}(\omega)$ is the residual magnetic field generated by the coils at the position of the magnetometer and $B_{z\chi}(\omega)$ represents the induced magnetic field in $z$ direction generated by test mass. $B_{zr}(\omega)$ must be suppressed to small enough compared with $B_{z\chi}(\omega)$ to increase the signal-to-noise ratio (SNR) for $B_{z\chi}(\omega)$ measurement.

In order to apply a spatially uniform magnetic field at the sample position while suppressing the background magnetic field created by the coils at the position of magnetometer, two concentric and coaxial Helmholtz coils, namely $C_1$ and $C_2$, are used. The radius of $C_1$ and $C_2$ coils are 15 cm and 5 cm, respectively. The origin of the coordinates is set to be the center of the Helmholtz coils and $z$ axis is set to be the axes of the coils. Currents with opposite directions are applied such that the magnetic field from $C_1$ and $C_2$ are opposite.

The magnetic field generated by each coil varies with the $z$ direction coordinate, as shown in Fig. 1(b). The cuboid sample is placed at the origin where the magnetic field is uniform and one side of the sample aligns with the $z$ axis. The difference in the radius of the coils ensure that at point $P$ far away from the origin, the magnetic field from $C_1$ and $C_2$ cancels each other and the total field $B_{zr}$ is nearly zero. By placing the fluxgate magnetometer at these positions, the SNR is significantly enhanced. The turns of $C_1$ and $C_2$ are designed to be 1000 and 3430, respectively. And the distance between point $P$ and the origin is calculated to be 13.6 cm, satisfying to the condition $D \gg L$.

The $C_1$ and $C_2$ coils are connected in series to suppress the noise from the current source. Within a 4 cm × 4 cm ×1 cm space, the uniformity of the magnetic field is about 94.72%, which is calculated by finite element simulation.

In the process of measuring $R(\omega)$ at different frequencies, an AC voltage with constant magnitude $V_{pp}$ is applied to the coils and the magnitude of current is $I_{pp} = V_{pp}/\sqrt{R_0^2 + \omega^2 L^2}$. The impedance of $C_1$ and $C_2$ coils connected in series is: $Z = R_0 + i\omega L$. With the rest of the coil parameters fixed, the position of the zero magnetic field point is determined by the number of turns of the $C_2$. More turns of $C_2$ pushes the zero magnetic field point, which is the position of the magnetometer, further away from sample. Therefore, in order to meet the requirement of $D \gg L$ and a strong enough magnetic field applied to the sample, the turns of $C_2$ need to be large.

Considering the geometry of the apparatus, all the coupling constants other than $K_{zz}$ in the matrix $\boldsymbol{K}$ is nearly to zero. As a result, the relationship between $B_{z\chi}$ and $M_{z\chi}$ can be simplified to $B_{z\chi} = K_{zz} M_{z\chi}$, giving $|R(\omega)| = \mu_0 B_{z\chi}/[K_{zz} B_z(\omega)]$. We then calculate $K_{zz}$ under two extremely different conditions. When the induced magnetic moment is uniformly distributed throughout the sample ($\omega \to 0$), $K_{zz}$ is calculated to be $7.805 \times 10^{-5}\ \mathrm{T/Am^2}$ assuming $I = 1$ A. On the other hand, when all the moment are concentrated on the surface of the sample because of the skin effect ($\omega \to \infty$), $K_{zz}$ is $7.845 \times 10^{-5}\ \mathrm{T/Am^2}$. The difference between the two $K_{zz}$ values is less than 0.5%, proving that when the distance between the test mass and the magnetometer is far enough, the measurement of $R(\omega)$ is insensitive to the particular distribution of the magnetic moment.

Another source of noise originates from the mechanical vibration of the magnetometer, which converts into a magnetic field noise by $\delta B_{gz} = (\partial B_z/\partial z)\delta z$ because there is a finite magnetic field gradient at the position of the magnetometer when the magnetic field is applied. To suppress the mechanical vibration noise, a rigid connection between the magnetometer and the coils was employed to minimize relative positional fluctuations.

Fig. 2 shows the spectrum of background magnetic field when the coils were energized and de-energized. The residual magnetic field was adjusted to be less than 4.0 nT for eliminating the contribution of current source noise to the assessment of $\delta B_{gz}$. The comparison of the two spectrum indicates that the magnetic noise caused by $\delta B_{gz}$ does not exceed 0.23 nT at 1 Hz. In order to reduce the magnetic field generated by magnetic components in the translation stage, the coils are mounted on a platform made of PMMA and wood, connected to the translation stage by a long wooden supporting pole. The position of the magnetometer is adjusted by the translation stage and the residual magnetic field can be adjusted to less than 1 ppm of the applied magnetic field at sample position.

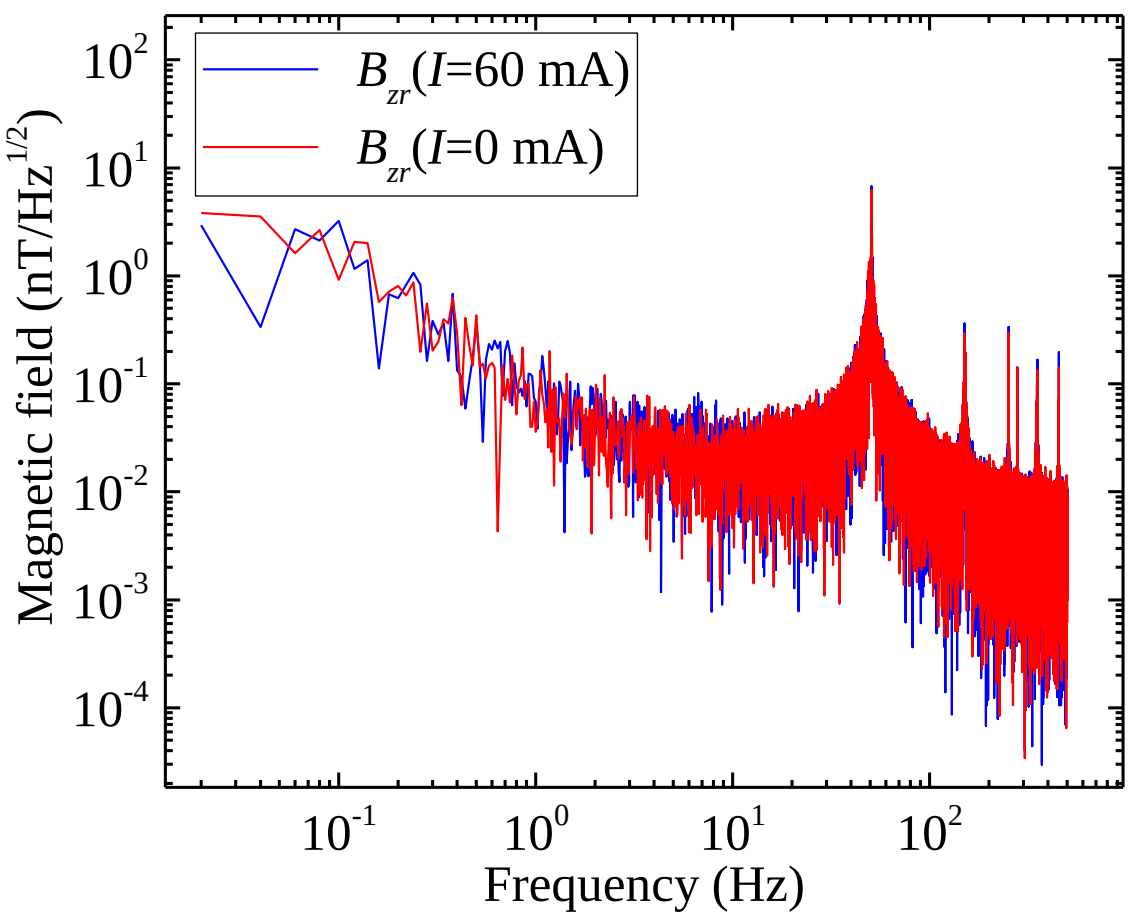

**Fig. 2.** Power spectral density of the background magnetic field. The red and blue lines are collected when the current in the coils $I = 0$ mA and $I = 60$ mA(DC), respectively.

The errors of the system for AC magnetic response measurement at 1 Hz and the combined uncertainty are presented in Table 1. Among them, the largest systematic error is the non-uniformity of the magnetic field, reaching about 5.28%. The total systematic error is 5.99%, and the uncertainty of repeated measurement for five times is 0.50%, which is considered as statistical error. The total error is taken as the synthetic error of systematic error and statistical error, which is about 6%. The measurement resolution of the induced magnetic field $\Delta B_{z\chi}$ is less than 0.3 nT, yielding a measurement resolution of $\Delta|R(\omega)|$ is about $1 \times 10^{-9}$ m$^3$ at 1 Hz. Since the maximum sampling rate of the fluxgate is 10000 Hz, the total measurement error increases to about 10% at 1500 Hz due to the insufficient sampling rate.

**Table 1.** The measurement error budget of the instrument at 1 Hz.

| Error terms | Error value | $\Delta\|R(\omega)\|/\ \|R(\omega)\|(\%)$ |
|---|---|---|
| Installation error of TM | 0.1 mm | 0.20 |
| Installation error of magnetometer | 0.1 mm | 0.01 |
| Installation error of coils | 0.1 mm | 0.01 |
| Non-uniformity of $B_z$ | 5.28% | $< 5.28$ |
| Non-uniformity of induced magnetic moment | / | $< 0.50$ |
| Geomagnetic field noise | $< 0.10$ nT/Hz$^{1/2}$ | $< 0.20$ |
| Background noise of magnetometer | $< 0.02$ nT/Hz$^{1/2}$ | $< 0.04$ |
| Current noise of coils | $10^{-3}$ | 0.01 |
| Magnetic field fluctuation caused by mechanical vibration | $< 0.23$ nT/Hz$^{1/2}$ | $< 0.46$ |

## III. DATA ANALYSIS

The magnetic field $B_{ztot}$ detected by the magnetometer with the sample placed on the sample holder can be expressed as：

$$B_{ztot} = B_{ztot_0}\cos(\omega t + \varphi_{ztot}) = B_{zr_0}\cos(\omega t + \varphi_{zr}) + B_{z\chi_0}\cos(\omega t + \varphi_{z\chi}) \quad (4)$$

Where $B_{zr} = B_{zr_0}\,cos(\omega t + \varphi_{zr})$ is the residual magnetic field generated by the Helmholtz coils, and $B_{z\chi} = B_{z\chi_0}\,cos(\omega t + \varphi_{z\chi})$ is the sample induced magnetic field.

Using the auxiliary angle formula, $B_{\chi z_0}$ and $tan\ (\varphi_{\chi z} - \varphi_z)$ can written in the form:

$$B_{z\chi_0} = \sqrt{{B_{zr_0}}^2 + {B_{ztot_0}}^2 - 2B_{zr_0}\,B_{ztot_0}\cos(\varphi_{ztot} - \varphi_{zr})} \quad (5)$$

$$\tan\left(\varphi_{z\chi}-\varphi_{zr}\right)=\frac{B_{ztot_0}\sin(\varphi_{ztot}-\varphi_{zr})}{B_{zr_0}+B_{ztot_0}\cos(\varphi_{ztot}-\varphi_{zr})} \tag{6}$$

In order to obtain the amplitude $B_{z\chi_0}$ and phase difference $\varphi_{z\chi}-\varphi_{zr}$, a phase reference signal with frequency $\omega$ is required. Since the magnetic fluxgate magnetometer is able to simultaneously measure the magnetic fields in all three directions, we can use the signal in the $y$ direction as phase reference. However, when the center of the test mass and magnetometer are both on the central axis shared by the two Helmholtz coils, the $y$ component of the residual magnetic field $B_{yr}=B_{yr_0}\,cos\left(\omega t+\varphi_{yr}\right)$ is zero. Therefore, we offset the magnetometer by 1 mm in the $y$ direction from the central axis for a finite $B_{yr}$, with $B_{yr}$ and applied field $B_z(\omega)$ in phase. In practice, deviations from the ideal conditions may occur due to the installation errors, resulting in a nonzero $B_{y\chi_0}$, contributing to the error of phase measurement. According to Eq. (6), the phase change in y direction caused by the induced magnetic field of the sample can be obtained as:

$$\delta\varphi_y=\left|\varphi_{ytot}-\varphi_{yr}\right|=\left|\arctan\left(\frac{B_{y\chi_0}\sin\left(\varphi_{y\chi}\right)}{B_{yr_0}+B_{y\chi_0}\cos\left(\varphi_{y\chi}\right)}\right)\right|=\left|\arctan(F)\right| \tag{7}$$

Taking $\varphi_{y\chi}$ as a variable, the upper limit of $F$ is obtained as:

$$F\le\frac{B_{y\chi_0}}{\sqrt{2}B_{yr_0}+B_{y\chi_0}}=\frac{1}{\sqrt{2}B_{yr_0}/B_{y\chi_0}+1} \tag{8}$$

Consequently, the maximum value of $\delta\varphi_y$ can be determined by Eq. (7) and Eq. (8). For Cu, the measured values of $B_{y\chi_0}$ is less than 0.7 nT at 1 Hz, which is much smaller than the value of $B_{yr_0}$=1600.0 nT, leading to only 0.03° in the change of $\varphi_{yr}$. $B_{y\chi_0}$ becomes larger when $|R(\omega)|$ is increased. At 1500 Hz, for Cu, the $y$ direction phase change $\delta\varphi_{yr}$ is less than 2.4°; for $Au_{0.7}Pt_{0.3}$, $\delta\varphi_{yr}$ is less than 1.0°. The phase of $R(\omega)$ above 1500 Hz is close to $-180°$, which means that the relative phase measurement error for the Au-Pt test mass is less than 1%.

The phase difference between $B_z$ and $B_y$ with and without the test mass is denoted as $\Delta\varphi_A = \varphi_{ztot} - \varphi_{yr}$ and $\Delta\varphi_B = \varphi_{zr} - \varphi_{yr}$, respectively, where $\varphi_{yr}$ does not change over time. Therefore, the phase difference $\varphi_{z_{tot}} - \varphi_{zr}$ is equal to $\Delta\varphi_A - \Delta\varphi_B$. $B_{z\chi_0}$ and $\varphi_{z\chi}$ can be solved by plugging the experimentally measured $B_{zr_0}$, $B_{ztot_0}$, $\varphi_{zr}$, $\varphi_{ztot} - \varphi_{zr}$ into Eq. (5) and Eq. (6). The modulus and phase of the magnetic response can be calculated as $|R(\omega)| = \mu_0 B_{z\chi_0}/[K_{zz}|B_{z_0}(\omega)|]$ and $\varphi(\omega) = \varphi_{z\chi}(\omega) - \varphi_{yr}(\omega)$.

## IV. RESULTS AND DISCUSSION

First, we measured $R(\omega)$ of a copper (Cu) cuboid sample of the size of $4\ \text{cm} \times 4\ \text{cm} \times 1\ \text{cm}$. Before performing measurements on the sample, the translation stage is adjusted to minimize the residual magnetic field $B_{zr}$, as depicted in Fig. 3.

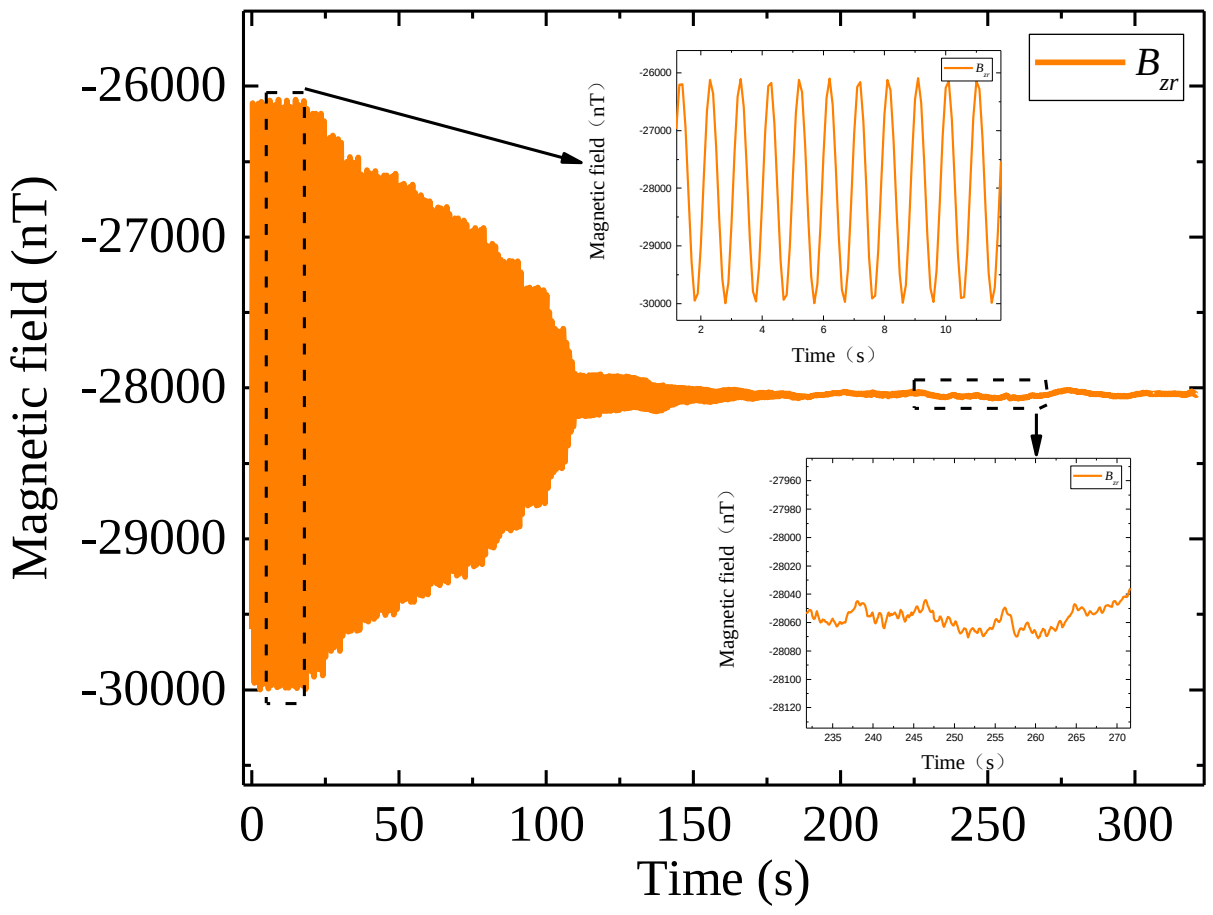

**Fig. 3.** The variation of residual magnetic field $B_{zr}$ when changing the position of the magnetometer. The residual magnetic field after the optimization is 2 nT at 1 Hz.

The time-domain variation curve of the magnetic field measured at frequencies of 1 Hz and 1500 Hz is shown in Fig. 4(a) and Fig. 4(b). The magnetic field changes significantly after the sample is placed on the holder. The fast Fourier transform (FFT) of the magnetic field is shown in the Fig. 4(c) and Fig .4(d), from which the magnetic field peak with the frequency of modulation can be obtained. To investigate whether AC magnetic response is related to conductivity, we also measured aluminum (Al) sample with the same size of Cu, which has a lower conductivity than Cu. The modulus and phase of $R(\omega)$ of both samples are displayed in Fig 5(a) and Fig 5(b).

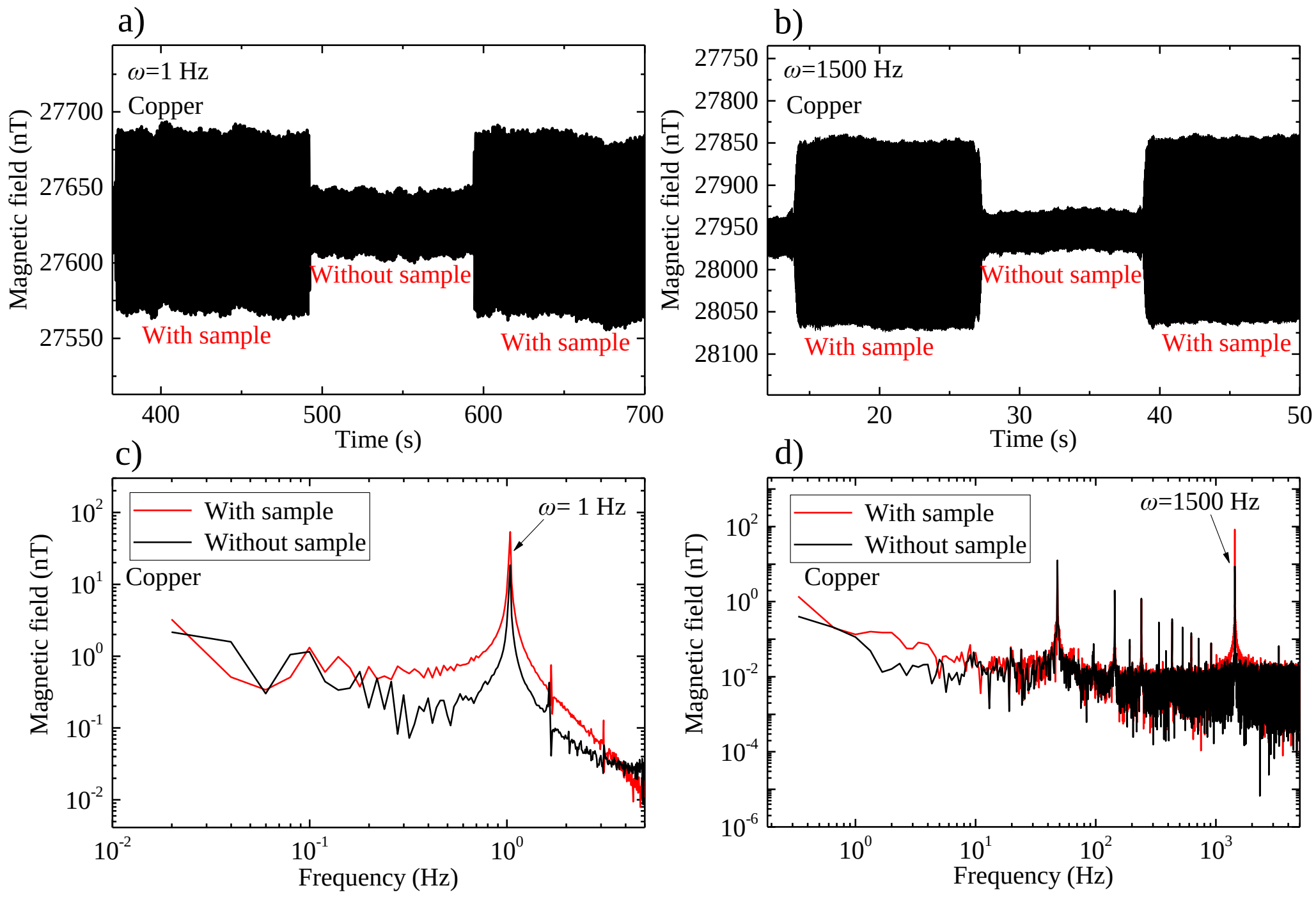

**Fig. 4. a) and b):** Magnetic field as a function of time measured at the frequency of 1 Hz and 1500 Hz respectively. **c) and d):** FFT of the measured time domain data.

According to the previous work, [27] the real part $R'(\omega)$ and imaginary part $R''(\omega)$ of the AC magnetic response $R(\omega)$ are related to the initial response $R_0$, frequency $\omega$, and relaxation time $\tau_e$, and are expressed as follows:

$$R'(\omega) = R_0 + A_0 \frac{\omega^2 \tau_e^2}{1 + \omega^2 \tau_e^2}, \qquad R''(\omega) = A_0 \frac{\omega \tau_e}{1 + \omega^2 \tau_e^2} \tag{9}$$

$$|R(\omega)| = \sqrt{R'^2 + R''^2}, \qquad \varphi(\omega) = \mathrm{atan}(R''/R') \tag{10}$$

Where the roll-of frequency $f_e$ represents the frequency at which the real and imaginary parts of $R(\omega)$ are equal, with relaxation time $\tau_e = 1/(2\pi f_e)$. Fitting the experiment data of $|R(\omega)|$ vs. frequency yield $f_e(\mathrm{Cu}) = (106 \pm 2)$ Hz and $A_0(\mathrm{Cu}) = -(10.40 \pm$ $0.14) \times 10^{-6}\ \mathrm{m}^3$. The value of $f_e(\mathrm{Cu})$ are consistent with the reference work, [16] confirming the reliability of our measurement method. Fitting the experimental data of $|R(\omega)|$ vs. frequency yielded $f_e(\mathrm{Al}) = (155 \pm 3)\ \mathrm{Hz}$ and $A_0(\mathrm{Al}) = -(9.10 \pm 0.11) \times 10^{-6}\ \mathrm{m}^3$.

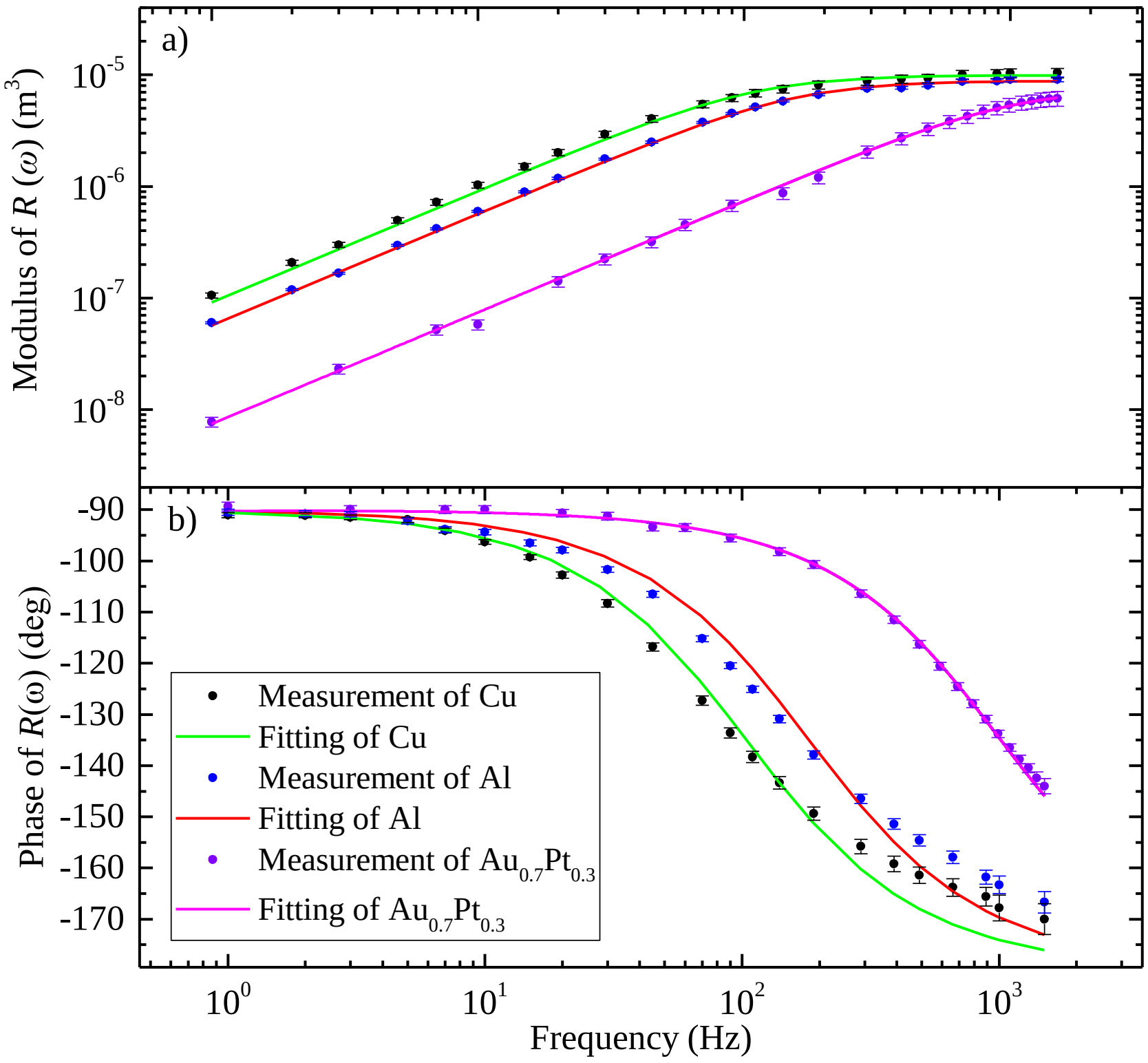

**Fig. 5. a):** AC magnetic response of modulus $|R(\omega)|$ of the cuboid samples as a function of frequency. **b):** AC magnetic response phase $\varphi(\omega)$ of the cuboid samples as a function of frequency. The solid lines are the fitting results using Eq. (9) and Eq. (10), the dots are the experimental measurement results.

For materials such as test mass, there are two magnetic field response mechanisms, namely quasi-static response and dynamic response. Due to the measurement frequency above 1 Hz, dynamic response dominates. The dynamic response is mainly caused by eddy currents. [27] The intensity of eddy currents is directly related to the material's conductivity $\sigma$, materials with higher conductivity exhibit more pronounced out-of-phase responses, leading to a greater phase lag in the magnetic response and subsequently longer relaxation time $\tau_e$. Cu has a higher conductivity than Al, therefore, Copper's $\tau_e$ is also larger. Furthermore, based on the measured relaxation time $\tau_e$ and known material conductivity $\sigma$, it was observed that $\tau_e(Cu)/\sigma(Cu) \approx \tau_e(Al)/\sigma(Al)$, suggesting a possible proportionality between relaxation time and conductivity. Despite the fact that there is a certain degree of relationship between conductivity and AC magnetic response, the conductivity test cannot be substituted for the AC magnetic response test. This is primarily because the conductivity is just one factor that contributes to the AC magnetic response of materials. Other factors, such as ferromagnetic impurities or magnetic domains, can also influence the magnetic response. By measuring AC magnetic response directly, we can better distinguish these effects and identify potential sources of error or noise and obtain a more complete picture of the test mass's AC magnetic properties. Therefore, it is necessary and indispensable to AC magnetic response measurement for evaluating the magnetic noise in space-borne gravitational wave detection.

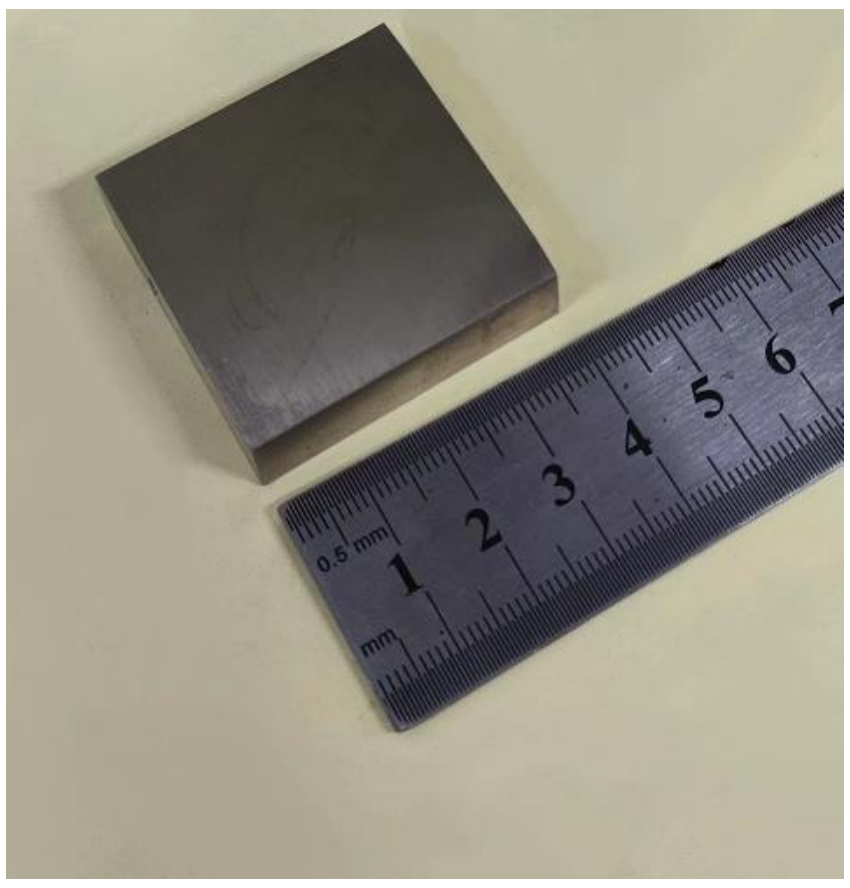

**Fig. 6**. The photograph of the $Au_{0.7}Pt_{0.3}$ alloy test mass.

In the gravitational wave detection mission, the test mass used is Au-Pt alloy with a elements ratio of Au to Pt approximately equal to 7:3. In order to verify whether our device can be used for measuring the material used in gravitational wave detection mission, we measured a $Au_{0.7}Pt_{0.3}$ alloy sample with a size of 4 cm×4 cm×1 cm, the physical image of the sample is shown in Fig. 6. The measurement results are shown in

Fig. 5. Fitting the experimental data of $|R(\omega)|$ vs. frequency yielded $f_e(\mathrm{Au_{0.7}Pt_{0.3}}) = (1015 \pm 10)$ Hz and $A_0(\mathrm{Au_{0.7}Pt_{0.3}}) = -(7.80 \pm 0.10) \times 10^{-6}\ \mathrm{m}^3$. It can be seen that compared to pure metals such as Cu and Al, $\mathrm{Au_{0.7}Pt_{0.3}}$ has a weaker response to AC magnetic field. It perhaps speculated that this is due to the decrease in conductivity caused by alloying process of Au and Pt.

## V. CONCLUSION

In conclusion, we developed an experimental apparatus to conveniently measure the AC magnetic response $R(\omega)$ of a cm-sized test mass for space-borne gravitational wave detection in ambient conditions with no magnetic shielding. The modulus and the phase of $R(\omega)$ are obtained by measuring the induced magnetic field while simultaneously monitoring the magnetic field in the orthogonal direction. The measurement frequency ranges from 1 Hz to 1500 Hz and can be extended to higher frequency by increasing the sampling rate of the magnetometer. This method is also applicable to other areas such as mineral research [28] and manufacturing [29] ,where the AC magnetic response measurement is required for large-size objects.

## ACKNOWLEDGMENTS

This work is supported by the National Key R&D Program of China (No. 2021YFC2202300), the Regional Innovation and Development Joint Fund of National Natural Science Foundation of China (No.U20A2077) , the Open Project of Yunnan Precious Metals Laboratory Co., Ltd (No.YPML-2023050249) and the National Key R&D Program of China (No.2023YFC2205801).